# Analytical approach for type-II semiconductor spherical core-shell quantum dots heterostructures with wide band gaps

Tiberius O. Cheche[1] and Yia-Chung Chang[2,a)]

[1]*Faculty of Physics, University of Bucharest, Bucharest 077125, Romania*

[2]*Research Center for Applied Sciences, Academia Sinica, Taipei 115, Taiwan*

A one-band model within the effective mass approximation is adopted to characterize the energy structure and oscillator strength of type-II semiconductor spherical core-shell quantum dots. The heteroepitaxial strain of the core-shell heterostructure is modeled by the elastic continuum approach. The model is applied to ZnTe/ZnSe core-shell, a wide band gap type-II heterostructure. The simulated absorption spectra are in fair agreement with available experimental results.

## I. Introduction

The atomic-like optical spectra of the semiconductor quantum dots (QDs) make these materials of considerable interest from the perspective of novel optical and electronic properties. The size and shape are factors of high-level control of the physical properties of such nanostructures. The colloidal core-shell (CS) QDs are heterostructures that can be obtained by chemical synthesis[1, 2, 3] with reproducible and controllable size and shape and low fabrication cost. Their high quantum yields and the possibility of tuning the band gap of the QD by the shell thickness have attracted many research groups. The type-II core-shell QDs have been shown to have useful application in solar cells due to the ease to separate electron hole pairs.[4] Thus, a simple yet realistic model for describing the energy structures will be highly desired for the design of these type-II core-shell QDs for device applications.

Theoretically, several approaches have been used to calculate the electronic structures of semiconductor QDs. For example, the tight-binding method[5], the effective bond-orbital model[6], the valence force field model[7] or first-principles calculations[8], the envelope-function

[a)] Author to whom correspondence should be addressed. Electronic mail: yiachang@gate.sinica.edu.tw

methods as the effective mass approximation[9] or multi-band approach[10, 11, 12] have been developed. Each of these methods has some limitations either in the accuracy of the predicted electronic structures or the computation cost.

In this work, we present a simple one-band model within the effective mass approximation to describe the energy structure and the optical properties of type-II CS QDs. The model is applied to materials where such an approach is expected to work well, namely in the case of: i) weak mixing of the conduction band (CB) and valence band (VB), and ii) weak mixing of the hole states within each compound of the heterostructure. The first condition is typically fulfilled by the wide band gap semiconductor heterostructures. As for the second, though the heavy hole-light hole admixture is a well known effect[11], in the first approximation, we may disregard it if considering small QDs, in which it was found that the admixture is less pronounced[11, 13, 14] due to the large energy splitting between the heavy- and light-hole subbands caused by the strong quantum confinement. The strain induced by the lattice mismatch at the interface is an important factor contributing to the band offsets, and we take it into account within an elastic continuum model. In the present calculations, we assume an ideal, defect–free crystal structure (technically the shell thickness should be of only several monolayers to fulfill this requirement). The analytical predictions of the model are compared with the experimental results reported for a type-II heterostructure that has wide band gap for the both compounds involved. As the model QD has spherical symmetry, the piezoelectric effect is expected to be less significant[15], and it is a good approximation to neglect it.

The structure of the paper is as follows. In Sec. 2 the theoretical model is introduced to describe the energy structure and oscillator strength. In Sec. 3 the model is applied to ZnTe/ZnSe (wide band gap heterostructure) CS QDs. The validity of the approximations used, and the comparison of the predicted and experimental absorption spectra are discussed in this section. Conclusions are given in Sec. 4. To make the work more comprehensible, we present details related to the strain computation in Appendix A, some considerations of the heavy and light states effective masses in the one-band spherical approximation in Appendix B, and detailed derivations of the oscillator strength in Appendix C.

## II. One-band model for core-shell semiconductor quantum dots

To obtain the energy structure within the one-band model one needs to know the bulk band-offset of the hetersotructure and the effective masses of the carriers. On the other hand, the strain induces a shift of the band edges and solving the problem of the strain influence on the band lineups in heterostructures is necessary. Thus, according to the *model-solid theory* developed by Van de Walle[16], both VB and CB are shifted by the hydrostatic deformation potential and, in addition, by the spin-orbit coupling in the case of the VB. The value of the band edge is given by[16]

$$E_v = E_v^0 + \frac{\Delta}{3} + a_v \frac{\Delta\Omega}{\Omega} \tag{1a}$$

$$E_c = E_c^0 + a_c \frac{\Delta\Omega}{\Omega} \tag{1b}$$

where $v$ ($c$) holds for VB (CB), $E_v^0$ and $E_c^0$ denote the unstrained values of the band edges, $\Delta$ is the spin-orbit splitting, $a_v$ and $a_c$ are the corresponding hydrostatic deformation potentials, $\Delta\Omega/\Omega = \varepsilon_{hyd}$ is the fractional volume change due to the hydrostatic strain.

For ZnTe/ZnSe CS QD, the type II heterostructure we discuss in this work, we draw in FIG. 1 the schematic band lineups corresponding to the strained case (notations are in accordance with those from Eqs. (1)). On the external surface of the shell, the potential is approximated by a hard wall.

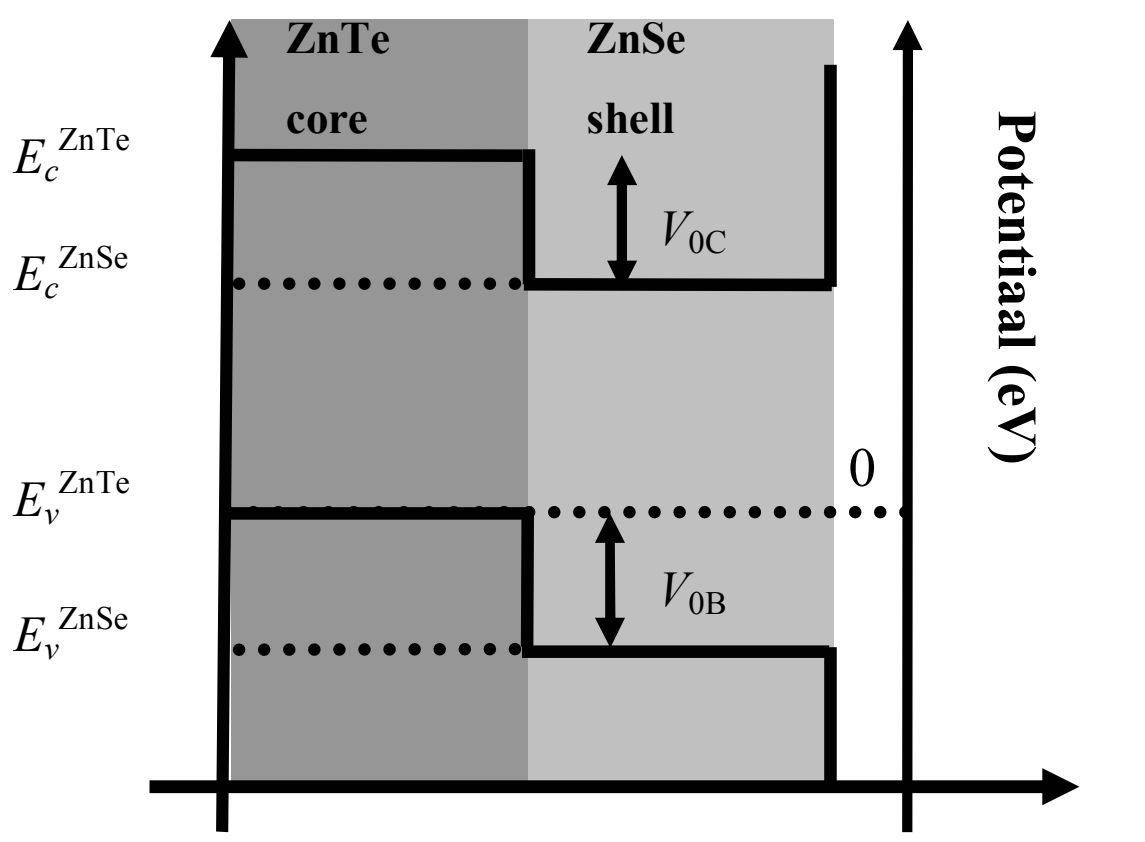

**FIG. 1** Schematic band lineups for CdTe/ZnSe core-shell QD in presence of strain. The notations of the energies are in accordance with Eq. (1).

According to FIG. 1, we consider the following spherical square-well potentials as functions of $r$:

i) for electrons:

$$V(r)=\begin{cases}V_{0\mathrm{C}},\ \text{if } 0\le r<r_0\\ 0,\ \text{if } r_0\le r<R\\ \infty,\ \text{if } r\ge R\end{cases};\tag{2a}$$

ii) for holes:

$$V(r)=\begin{cases}0,\ \text{if } 0\le r<r_0\\ V_{0\mathrm{B}},\ \text{if } r_0\le r<R\\ \infty,\ \text{if } r\ge R\end{cases};\tag{2b}$$

with $V_{0\mathrm{C}}=E_c^{\mathrm{ZnTe}}-E_c^{\mathrm{ZnSe}}>0$, $V_{0\mathrm{B}}=-E_v^{\mathrm{ZnSe}}>0$.

The one-particle radial Schrödinger equation for the spherical well potential $V(r)$ given by Eqs. (2a, b) reads

$$\frac{-\hbar^2}{2m^*(r)}\frac{1}{r}\frac{d^2}{dr^2}\left[rR_l(r)\right]+\frac{l(l+1)\hbar^2}{2m^*(r)r^2}R_l(r)=\left[E-V(r)\right]R_l(r),\tag{3a}$$

where $m^*(r)$ is the position-dependent effective mass. Using the change of variable, $\rho=k_ir$, where $k_i^2=\mp 2m_i(E-V_i)>0$ if $E\lessgtr V_i$ (with the notations, $m_i=m_{1,2}^{e,h}$ for the effective mass of the electron ($e$) or hole ($h$) in the core ($i$=1) or shell ($i$=2), and $V_i$ for the values of the piecewise-constant potential in region $i$, that is, $V_i$= 0 or $V_{0\mathrm{B}}$ or $V_{0\mathrm{C}}$), Eq. (3a) reduces to the Bessel differential equation

$$\rho^2\frac{d^2v_l(\rho)}{d\rho^2}+2\rho\frac{dv_l(\rho)}{d\rho}-l(l+1)v_l(\rho)\mp\rho^2v_l(\rho)=0,\tag{3b}$$

and consequently, the general solution of Eq. (3a) is of the form $R_l(r)=v_l(k_ir)$. In Eq. (3b) the sign of the last term is positive for $E>V_i$ and negative for $E<V_i$. The solutions of Eq. (3b) are linear combinations of spherical Bessel or modified spherical Bessel functions, namely: $v_l(\rho)=A_lj_l(\rho)+B_ly_l(\rho)$ if $E>V_i$ or $v_l(\rho)=A_li_l(\rho)+B_lk_l(\rho)$ if $E<V_i$, where $j_l(\rho), y_l(\rho)$ are the spherical Bessel functions of the first and second kind, and $i_l(\rho), k_l(\rho)$ are the modified spherical Bessel functions of the first and second kind, respectively. The solutions should be regular and consequently they should be of the following form.

*Electronic states*

For $0\le r\le r_0$:

$$E<V_{0C},\ R_l^{(1e)}(r)=A_l^ei_l(k_1^er),\tag{4a}$$

$$E>V_{0C},\ R_l^{(2e)}(r)=B_l^ej_l(k_2^er).\tag{4b}$$

For $r_0 \le r \le R$:

$$R_l^{(3e)}(r) = \left[C_l^e j_l(k_3^e r) + D_l^e y_l(k_3^e r)\right]. \tag{4c}$$

In Eqs. (4), $k_1^e = \sqrt{2m_1^e(-E+V_{0C})}\big/\hbar$ , $k_2^e = \sqrt{2m_1^e(E-V_{0C})}\big/\hbar$ , $k_3^e = \sqrt{2m_2^e E}\big/\hbar$ , and $A_l^e$, $B_l^e$, $C_l^h$, $D_l^h$ are orthonormalization constants.

The explicit expressions of the wave functions are obtained by imposing the following boundary conditions: the wave functions and probability current are continuous at the interfaces, and the wave functions vanish outside the QD.

For the case $E < V_{0C}$:

$$\begin{cases} R_l^{(1e)}(r_0) = R_l^{(3e)}(r_0) \\ \dfrac{1}{m_1^e}\dfrac{dR_l^{(1e)}(r)}{dr}\bigg|_{r_0} = \dfrac{1}{m_2^e}\dfrac{dR_l^{(3e)}(r)}{dr}\bigg|_{r_0} \\ R_l^{(3e)}(R) = 0 \end{cases}, \tag{5a}$$

Then, the energy is obtained by solving the transcendental equation

$$\frac{m_2^e}{m_1^e}\frac{i_l'(k_1^e r_0)}{i_l(k_1^e r_0)} = \frac{j_l'(k_3^e r_0) - j_l(k_3^e R)\, y_l'(k_3^e r_0)\big/ y_l(k_3^e R)}{j_l(k_3^e r_0) - j_l(k_3^e R)\, y_l(k_3^e r_0)\big/ y_l(k_3^e R)}. \tag{5b}$$

For the case $E > V_{0C}$ the equations expressing the boundary conditions and solutions are similar with those from Eq. (5), by replacing $R_l^{(1e)}$ by $R_l^{(2e)}$, $k_1^e$ by $k_2^e$, and $i_l$ by $j_l$.

*Hole states*

For $0 \le r \le r_0$:

$$R_l^{(1h)}(r) = A_l^h j_l(k_1^h r). \tag{6a}$$

For $r_0 \le r \le R$:

$$E < V_{0\mathrm{B}},\ R_l^{(2h)}(r) = \left[B_l^h k_l(k_2^h r) + F_l^h i_l(k_2^h r)\right], \tag{6b}$$

$$E > V_{0\mathrm{B}},\ R_l^{(3h)}(r) = \left[C_l^h j_l(k_3^h r) + D_l^h y_l(k_3^h r)\right]. \tag{6c}$$

In Eqs. (6), $k_1^h = \sqrt{2m_1^h E}\big/\hbar$ , $k_2^h = \sqrt{2m_2^h(-E+V_{0B})}\big/\hbar$ , $k_3^h = \sqrt{2m_2^h(E-V_{0B})}\big/\hbar$ , and $A_l^h$, $B_l^h$, $F_l^h$, $C_l^h$, $D_l^h$ are orthonormalization constants.

The explicit expressions of the wave functions are obtained by imposing the following conditions for continuity of the wave functions and probability current at the interfaces.

Case $E < V_{0B}$:

$$\begin{cases} R_l^{(1h)}(r_0) = R_l^{(2h)}(r_0) \\ \dfrac{1}{m_1^h}\dfrac{dR_l^{(1h)}(r)}{dr}\bigg|_{r_0} = \dfrac{1}{m_2^h}\dfrac{dR_l^{(2h)}(r)}{dr}\bigg|_{r_0} \\ R_l^{(2h)}(R) = 0 \end{cases}, \tag{7a}$$

Then, the energy is obtained by solving the transcendental equation

$$\frac{m_2^h}{m_1^h}\frac{j_l'(k_1^h r_0)}{j_l(k_1^h r_0)} = \frac{-k_l'(k_2^h r_0) + k_l(k_2^h R) i_l'(k_2^h r_0)/i_l(k_2^h R)}{-k_l(k_2^h r_0) + k_l(k_2^h R) i_l(k_2^h r_0)/i_l(k_2^h R)}. \tag{7b}$$

For the case $E > V_{0B}$ the equations expressing the boundary conditions and solutions are similar with those from Eqs. (7), by replacing $R_l^{(2h)}$ by $R_l^{(3h)}$, $k_2^h$ by $k_3^h$, $i_l$ by $y_l$, and $k_l$ by $j_l$. The conditions $R_l^{(3e)}(R) = 0$ and $R_l^{(2h,3h)}(R) = 0$ impose the wave functions have a node on the external surface of the shell.

The solutions for the energy and normalizations constants of Eqs. (5) and (7) implies knowing of the strain. The implementation of strain effect is considered as follows. Within the one-band model the total wave function of the QD is given by the product of the envelope wave function and cell Bloch wave function at a certain point in ***k***-space of minimum energy (the $\Gamma$ point for the structure analyzed in Sec. 3). In terms of state vectors, we have $|\Psi_{nLm\alpha}\rangle = |\psi_{nLm}\rangle |u_\alpha\rangle$ ; $|\Psi_{nLm\alpha}\rangle$, $|\psi_{nLm}\rangle$, and $|u_\alpha\rangle$ denote the total, envelope, and the Bloch state vector at the band edge, respectively, and $\alpha = v, c$ holds for VB, CB, respectively, and the index *n* labels the energy level in ascending order for a given *l*. In Sec. 3, we consider the application of this model to a II-VI zincblende heterostructure. In both compounds of this type, the heavy and light hole states are degenerate at the $\Gamma$ point[17]. On the other hand, the spin-orbit interaction and shear strain can induce band splitting, as we already mentioned at the beginning of this section. The first splitting caused by the spin-orbit interaction raises the light and heavy hole band edge energy with respect to the split-off hole band edge as described by Eq. (1a). The second splitting, which is caused by the shear strain is neglected in this work and an argument for this is given in Appendix A. In addition, as we mentioned in Introduction, we consider the heavy and light hole states as unmixed (and both far from the split-off bands). Consequently, we may apply Eq. (7) to the heavy and light holes separately, by using the same $V_{0\mathrm{B}}$, but different effective masses. In the calculation, we consider the spherical part of the heavy-hole masss ($m^{hh}$) and light-hole masses ($m^{lh}$) assumed by the parabolic dispersion of the one-band model (see Appendix B),

$$m^{hh,(lh)} = \frac{m_0}{\gamma_1}\left[1-(+)\frac{6\gamma_3+4\gamma_2}{5\gamma_1}\right] \tag{8}$$

Finally, by superposing the two sets of hole states, we obtain the approximate VB energy structure.

Thus, concluding the discussion regarding the strain effect, to obtain the energy structure also requires knowing the hydrostatic strain. For the core-shell geometry, within the continuum elasticity approach, we obtain (see Appendix A):

$$\varepsilon_{1hyd} = 2\varepsilon_0\left[1-\left(\frac{r_0}{R}\right)^3\right]\frac{1-2\nu_1}{1-\nu_1}, \tag{9a}$$

$$\varepsilon_{2hyd} = -\frac{2\varepsilon_0}{3}\left(\frac{r_0}{R}\right)^3\frac{1-2\nu_2}{1-\nu_2}, \tag{9b}$$

where the subscript 1(2) holds for the core (shell), $\varepsilon_0$ is the relative mismatch, and $\nu_{1,(2)}$ is the Poisson ratio of the core (shell) material. For ZnTe/ZnSe CS QD, the case we analyze in section 3, the lattice constant of ZnTe, $a_1$, is larger than the lattice constant of ZnSe, $a_2$, and one obtains the strain is compressive for the core and tensile for the shell. Quantitatively, the relative mismatch $\varepsilon_0 = (a_2 - a_1)/a_1$ is negative and according to Eqs. (9), $\varepsilon_{1hyd} < 0$ (compression), and $\varepsilon_{2hyd} > 0$ (dilation).

## III. Application to ZnTe/ZnSe core-shell quantum dots

The band lineups in presence of strain are obtained by using the model-solid theory of Van de Walle [16] as follows. For the bulk (unstrained) band-offset, we consider the gap energies [18], $E_{g0}^{\mathrm{ZnTe}} = 2.25\mathrm{eV}$, $E_{g0}^{\mathrm{ZnSe}} = 2.69\mathrm{V}$, and $a_1 = 0.6103\mathrm{nm}$, $a_2 = 0.5668\mathrm{nm}$, $v_1 = 0.363\mathrm{nm}$, $v_2 = 0.375\mathrm{nm}$, $\Delta_{\mathrm{ZnTe}} = 0.91\mathrm{eV}$, $\Delta_{\mathrm{ZnSe}} = 0.43\mathrm{eV}$. Then, from Ref. 16, $E_{v0}^{\mathrm{ZnTe}} = -7.17\mathrm{eV}$, $E_{v0}^{\mathrm{ZnSe}} = -8.37\mathrm{eV}$, $E_{c0}^{\mathrm{ZnTe}} = E_{g0}^{\mathrm{ZnTe}} + E_{v0}^{\mathrm{ZnTe}} + \Delta_{\mathrm{ZnTe}}/3 = -4.62\mathrm{eV}$, and $E_{c0}^{\mathrm{ZnSe}} = E_{g0}^{\mathrm{ZnSe}} + E_{v0}^{\mathrm{ZnSe}} + \Delta_{\mathrm{ZnSe}}/3 = -5.54\mathrm{eV}$. With these bulk values of the energies and by using Eqs. (1) and (9) (with $\varepsilon_0 = -0.071$), we find the band lineups in presence of strain as function of $R$ and $r_0$ as shown in FIG. 2. One obtains that the strain induces enlargement (shrinkage) of the band gap for ZnTe (ZnSe), and the band gaps, $E_g^{\mathrm{ZnTe}} = E_c^{\mathrm{ZnTe}} - E_v^{\mathrm{ZnTe}}$, $E_g^{\mathrm{ZnSe}} = E_c^{\mathrm{ZnSe}} - E_v^{\mathrm{ZnSe}}$, $E_g = E_c^{\mathrm{ZnSe}} - E_v^{\mathrm{ZnTe}}$ increase with the shell thickness.

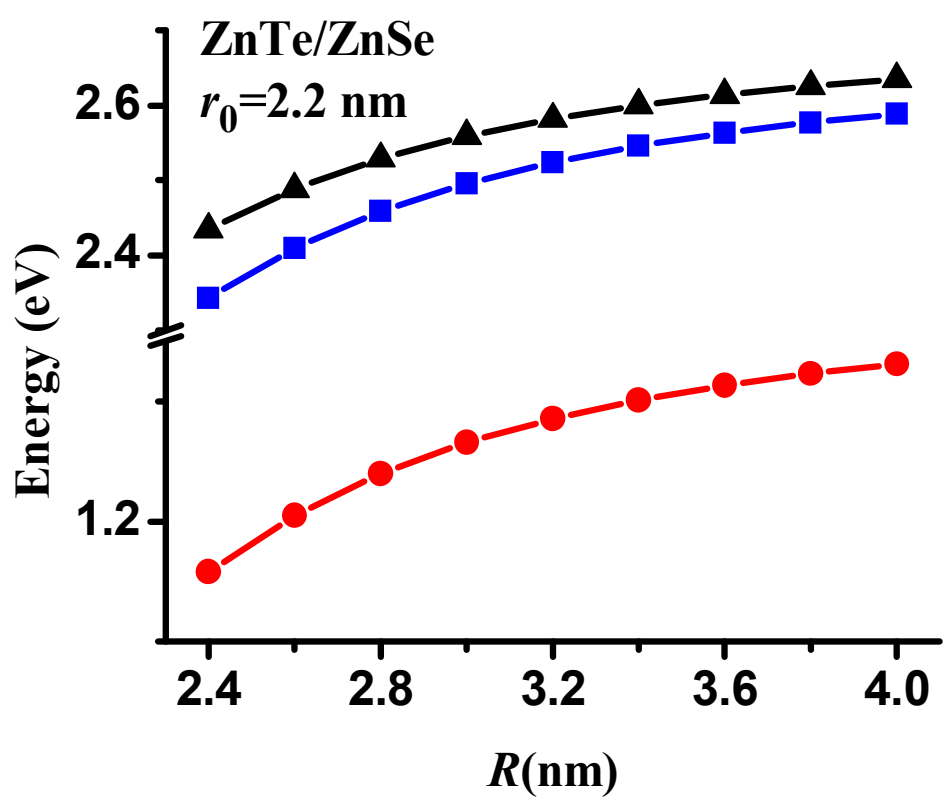

**FIG. 2** Gap energies of ZnTe/ZnSe CS QDs in the presence of lattice mismatch strain vs. core+shell radius $R$, with the core radius $r_0 = 2.2\text{nm}$. The legend:

■ for $E_g^{\text{ZnTe}} = E_c^{\text{ZnTe}} - E_v^{\text{ZnTe}}$;

▲ for $E_g^{\text{ZnSe}} = E_c^{\text{ZnSe}} - E_v^{\text{ZnSe}}$;

● for $E_g = E_c^{\text{ZnSe}} - E_v^{\text{ZnTe}}$.

Then, with the values of the Luttinger parameters from Lawaetz[19], $\gamma_1^{ZnTe} = 3.74$, $\gamma_2^{ZnTe} = 1.07$, $\gamma_3^{ZnTe} = 1.64$, $\gamma_1^{ZnSe} = 3.74$, $\gamma_2^{ZnSe} = 1.24$, $\gamma_3^{ZnSe} = 1.67$ the spherically averaged hole effective masses according to Eq. (8) are $m_{\text{ZnTe}}^{hh} = 1.09m_0$, $m_{\text{ZnTe}}^{lh} = 0.15m_0$, $m_{\text{ZnSe}}^{hh} = 1.29m_0$, $m_{\text{ZnSe}}^{lh} = 0.15m_0$. For electrons, we consider the effective masses, $m_{\text{ZnTe}} = 0.20m_0$, and $m_{\text{ZnSe}} = 0.21m_0$[20]. The energy structure is shown in FIG. 3 for the first four levels for both electrons and holes states. With Eq. (7), when comparing the energies values obtained with heavy and light holes, we obtain the first four hole states are heavy hole states. One can see the hole energies in FIG. 3 remain practically not affected by the shell thickness when the mixing of the hole states of the two compounds is absent (large $V_{0\text{B}}$ of the heterostructure). On the other hand, electron energy decreases with the shell thickness, which results in decreasing of the lowest energy transition, in accordance with the results regarding the absorption and emission spectra reported by Bang *et al.*[21] for such ZnTe/ZnSe CS QDs.

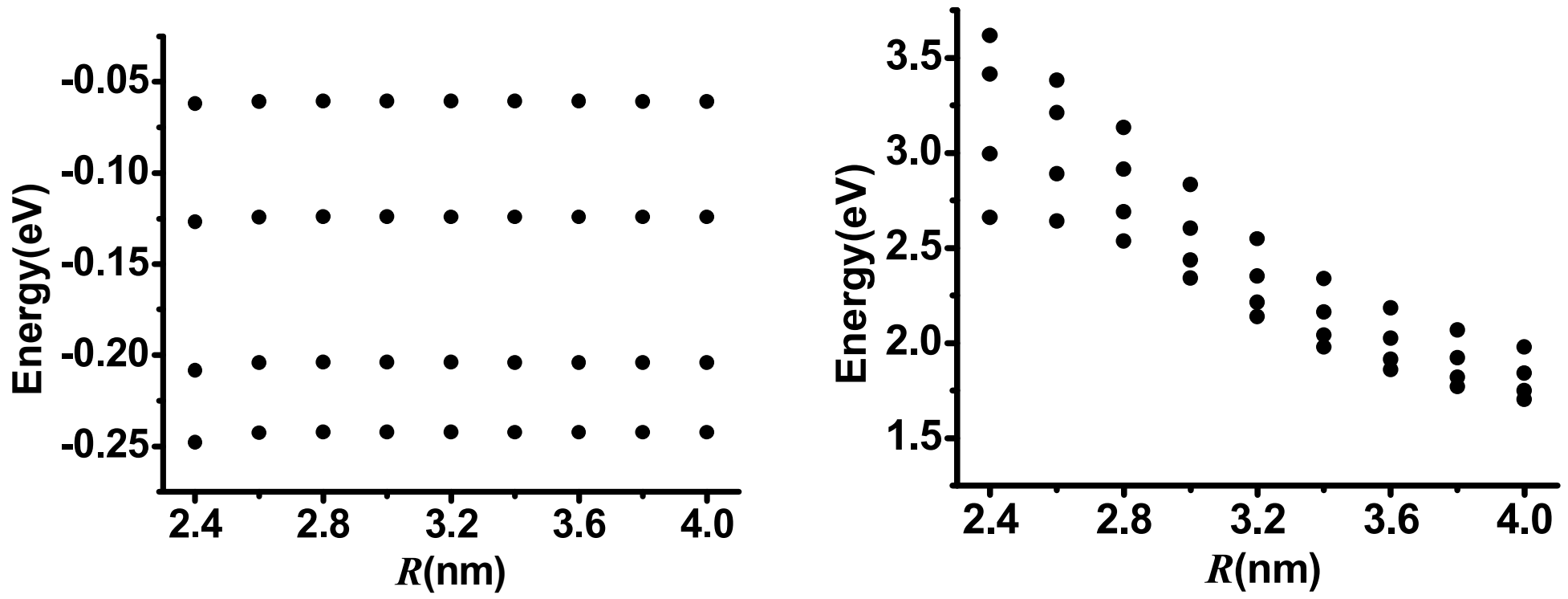

(a) (b)

**FIG. 3** Energy levels of: (a) holes; (b) electrons for ZnTe/ZnSe CS QDs vs. core+shell radius *R*, with the core radius $r_0 = 2.2\mathrm{nm}$.

Next, we present the results obtained from our model. Though the hole states mixing influences the shape of the orbitals as a result of mixing of different types of orbitals (*s*, *p*, *d*, …) in the exact form of the envelope wave function, we expect for this heterostucture with wide band gap, as a result of weak mixing of the hole states, the exact orbitals and those obtained by this one-band model to have similar features. The envelope wave function is obtained as the product of the radial wave function and the spherical harmonics $Y_{lm}(\theta, \varphi)$. For axis *z chosen* as quantization axis, the spatial density probabilities, the orbitals, $\left|R_{nl}^{(\alpha)}(r)Y_{lm}(\theta, \varphi)\right|^2$ are represented in FIG. 4 for $R = 2.4, 3, 3.6(\mathrm{nm})$ and $r_0 = 2.2\mathrm{nm}$. One can see transition from type-I (both electron and hole localized in the core) to type-II heterostructure, and shrinking (stretching) of the hole (electron) orbitals with shell thickness. For thin shell, the electron is still localized in the core while for thicker ones the electrons migrates to the shell ('falls down into the trap' opened by the potential of the shell, see FIG. 1). The hole remains confined in the core. The ground state has spherical symmetry for both electron and hole, which is characteristic for the *s* orbital.

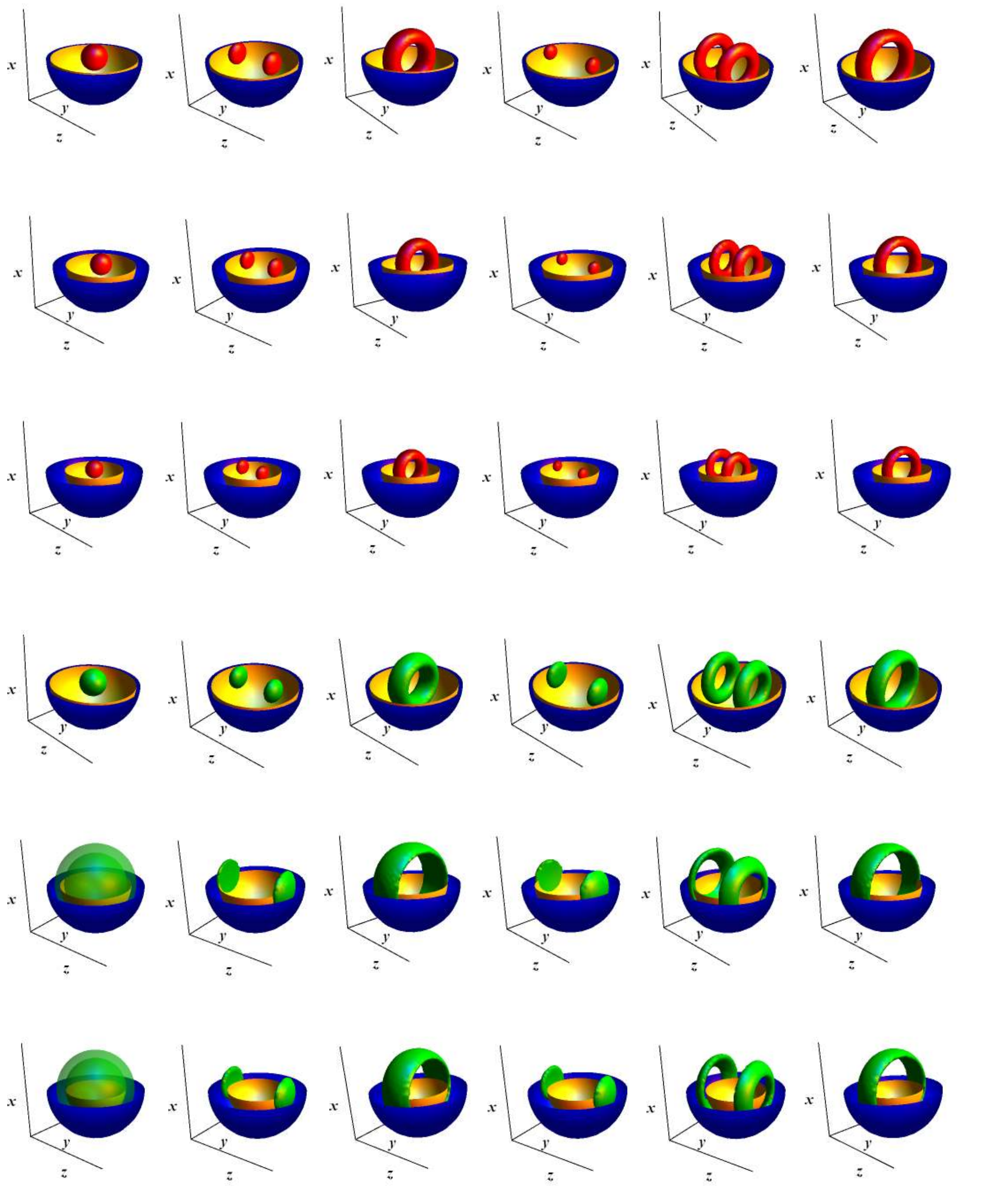

(a) (b) (c) (d) (e) (f)

**FIG. 4.** Hole (red color) and electron (green color) probability density $\mathrm{Max}(\left|\Psi_{nlm}^{h,e}(\mathbf{r})\right|^2)\times 0.75$ for ZnTe/ZnSe CS QD with core radius $r_0 = 2.2\mathrm{nm}$: the first three rows are for holes and the next for electrons for the radius $R = 2.4,\ 3,\ 3.6(\mathrm{nm})$. The figures are denoted according to the following quantum numbers: (a) $n=1,\ l=0,\ m=0$; (b) $n=1,\ l=1,\ m=0$; (c) $n=1,\ l=1,\ m=\pm 1$; (d) $n=1,\ l=2,\ m=0$; (e) $n=1,\ l=2,\ m=\pm 1$; (f) $n=1,\ l=2,\ m=\pm 2$. The orbitals correspond to the first three electron (hole) energy levels in ascending (descending) order from Fig. 3 ((a) to the first, (b) and (c) to the second, and (d)-(f) to the third).

The optical spectra can be described by the oscillator strength (see Appendix C)

$$f_{ij} = \frac{E_P}{\hbar\omega_{ij}} \left| \left\langle \psi^*_{nLm} \middle| \psi_{n'L'm'} \right\rangle \right|^2 , \tag{10}$$

which characterizes the probability of interband transition between two states, $i$ (characterized by the set of quantum numbers, $n, L, m$ ) and $j$ (characterized by the set of quantum numbers, $n', L', m'$ ); $E_P = 2m_0 |P|^2 / \hbar^2$ and $P = -i(\hbar / m_0)\langle s | p_x | x \rangle$ is the Kane momentum matrix element. In FIG. 5 we present the influence of the shell thickness on the oscillator strength obtained from the first four hole and electron states described in FIGs. 3 and 4 for $E_P = 19.1\text{eV}$ [13]. It follows from the overlap of the envelope wave functions, which is larger for thinner shell (type-I character of the heterostructure) as the orbitals from FIG. 4 show it. Thus, thicker shell leads to a decrease of the oscillator strength and consequently, a possible reduction of the quantum yield of the QD as reported by Bang et al.[21] for the same kind of QDs. Also a continuous red shift of the lowest energy transition, **1** (as denoted in FIG. 5) with the shell thickness is obtained, similarly to the experiment[21].

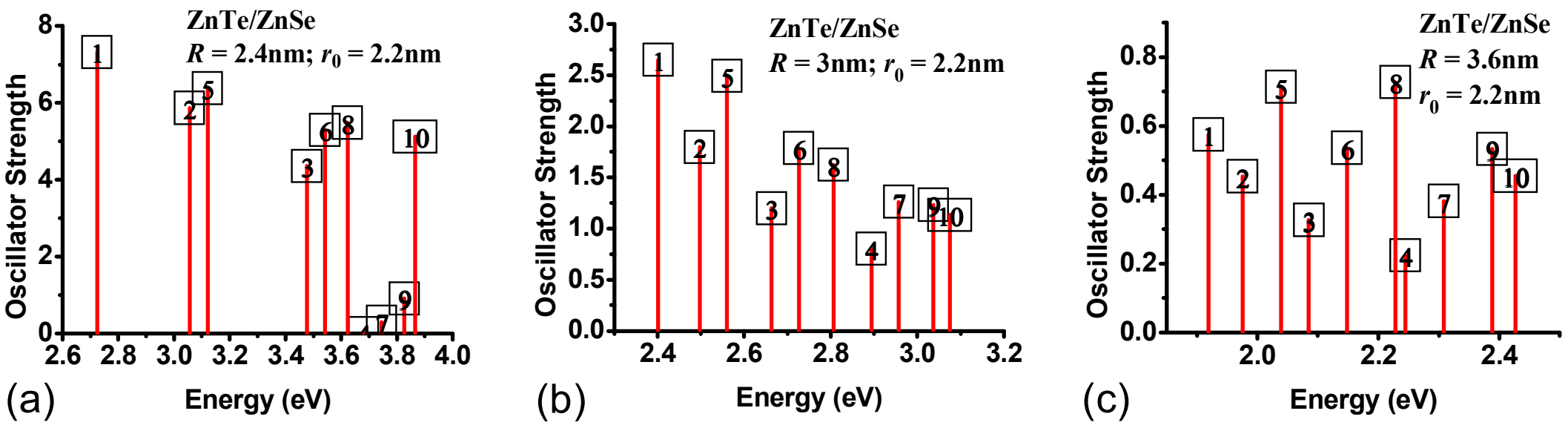

**FIG. 5** The oscillator strength for ZnTe/ZnSe CS QD with $r_0 = 2.2\text{nm}$ and a) $R = 2.4\text{nm}$, b) $R = 3\text{nm}$ , c) $R = 3.6\text{nm}$ . The labels in the figures correspond to the following transitions: **1** for $1 \leftrightarrow 1$, **2** for $1 \leftrightarrow 2$, **3** for $1 \leftrightarrow 3$, **4** for $1 \leftrightarrow 4$, **5** for $2 \leftrightarrow 2$, etc. The labels *1*, *2*, *3*, *4*, represent the four electron (holes) states from Fig. 3 in ascending (descending) order of the energy.

In FIG. 6, we compare the absorption results obtained by Bang et al.[21] with our simulated results, for ZnTe/ZnSe CS QDs with $r_0 = 2.2\text{nm}$ as function of shell thickness. We assign the lowest energy transition, **1**, active one as shown in FIG. 5 to the main peak recorded in Ref. 21 in absorbance measurements. The experimental data are adapted from the absorbance in FIG. 2 (a) from Ref. 21, by taking the ZnSe monolayer thickness of 2.83Å[22]. Experimentally, obtaining uniformly coated cores with spherical shape for the shell is difficult task, but comparison between frequencies of the absorbed light provides a reasonable fit for the simplicity of the model we used.

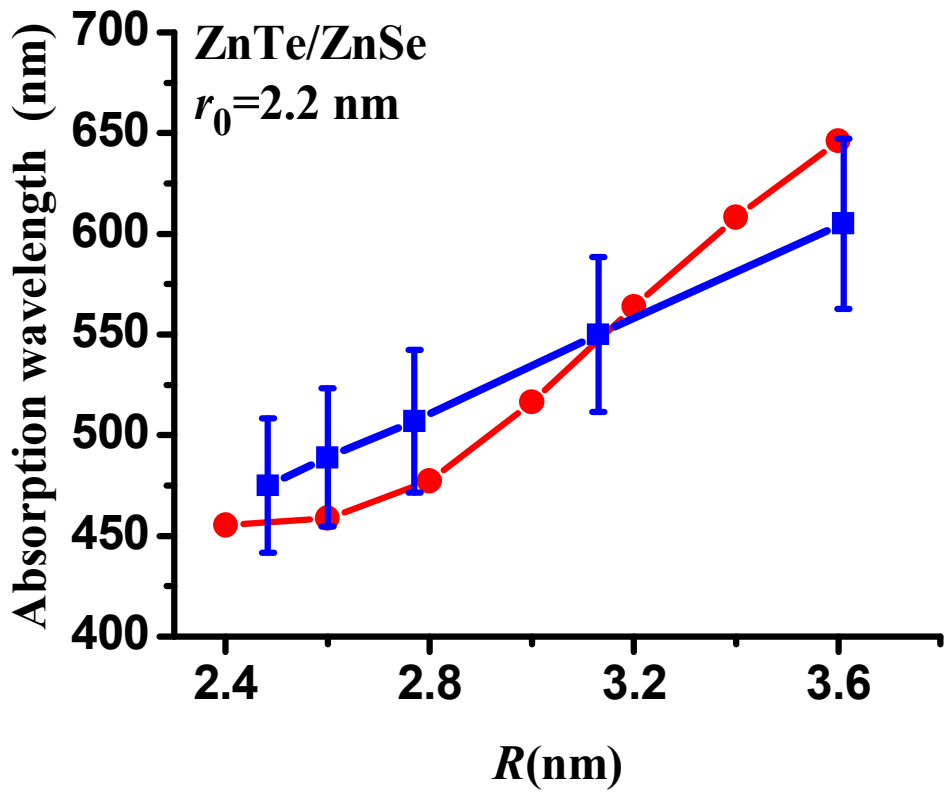

**FIG. 6** Absorption wavelength for ZnTe/ZnSe CS QDs with $r_0 = 2.2\text{nm}$ as a function of the shell thickness. The legend: ■ - experimental values and ● - simulated values. The simulation is obtained for the lowest energy transition corresponding to the transition **1** (*1* ↔ *1*) in Fig. **5**. The error bars of the experimental values are of 7%.

## IV. Conclusions

The one-band model within effective mass approximation is adopted to explain the energy structure of CS QDs, giving reasonably good results, as expected for the wide band gap ZnTe/ZnSe heterostructures. The use of the heavy hole effective mass in calculation is justified and the rather large effective masses obtained within the spherical approach, $m_{\text{ZnTe}}^{hh} = 1.09m_0$, $m_{\text{ZnSe}}^{hh} = 1.29m_0$ are important for a closer fit of the experimental data of absorption. The excitonic effect is expected to lead to a correction to the results. We can get approximately the influence of the electron-hole Coulomb interaction in spherical nanostructures by modeling this interaction with the expression, $-1.8e^2/(4\pi\varepsilon_{vac}\varepsilon R)$,[23] where $\varepsilon$ is the relative dielectric constant at high frequency. With $\varepsilon = 7.3$ [24] as the value for a homogenous ZnTe core of radius $R$, the Coulombic interaction has values in the interval $0.09 \sim 0.14\text{eV}$ for $R = 2.4 \sim 4\text{nm}$, and the simulated absorption from FIG. 6 is displaced upward with an excellent fit for the interval $R = 2.4 \sim 3.1\text{nm}$. On the other hand, introduction of the strain is absolutely necessary for getting a reasonable level of accuracy of the predictions, and expressions of the hydrostatic strain (easy to be implemented in the calculation) are obtained. Given its simplicity, robustness, and satisfactory accuracy level, we think the model we presented is very useful for at least preliminary calculations of the optical properties of type-II CS QDs of wide band gap, where the band mixing effect is of less importance.

## Acknowledgments

This work was supported in part by Academia Sinica and National Science Council, Taiwan under Contract NSC 101-2112-M-001-024-MY3, the grant POSDRU/89/1.5/S/58852, Project ‘‘Postdoctoral programme for training scientific researcher’’ co-financed by the European Social Found within the Sectorial Operational Program Human Resources Development 2007-2013, and the grant of the Romanian National Authority for Scientific Research, CNCS – UEFISCDI, project number PN-II-ID-PCE-2011-3-1007.

## Appendix A: Strain calculation

The strain tensor for CS QDs is derived starting from the radial displacement in a hollow sphere with inner and outer radius $r_0$ and *R*, which is subjected to an inner and outer pressure $P_{int}$ and $P_{out}$ [25, 26], respectively, that reads,

$$u(r,r_{int},r_{out},P_{int},P_{out}) = \frac{\pi\,(1+v)r}{E}\left( P_{int}\,\frac{\dfrac{1-2v}{1+v}+\dfrac{1}{2}\left(\dfrac{r_{out}}{r}\right)^3}{\left(\dfrac{r}{r_{int}}\right)^3-1} + P_{out}\,\frac{\dfrac{1-2v}{1+v}+\dfrac{1}{2}\left(\dfrac{r_{int}}{r}\right)^3}{\left(\dfrac{r_{int}}{r_{out}}\right)^3-1} \right). \tag{A1}$$

For the core

$$u_{int}(r,P) = u(r,r_{out},r_{int},P_{int},P_{out})\Big|_{r_{int}=0;\; r_{out}=r_0\;\; P_{int}=0;\; P_{out}=P}. \tag{A2}$$

For the shell

$$u_{out}(r,P) = u(r,r_{out},r_{int},P_{int},P_e)\Big|_{r_{out}=R;\; r_{int}=r_0;\; P_{int}=P;\; P_{out}=0}. \tag{A3}$$

For pseudomorphic growth, the relative mismatch of the two materials is $\varepsilon_0 = (a_2 - a_1)/a_1$ (with $a_1$ the core lattice constant, and $a_2$ the shell lattice constant). The inner and outer material impose the *shrink–fit* condition

$$u_{int}(r_0, P) - u_{out}(r_0, P) = \varepsilon_0 r_0 \ . \tag{A4}$$

Thus, with Eqs. (A1-A3) we obtain

$$u_{int}(r) = \frac{2}{3}\varepsilon_0 r\left[1-\left(\frac{r_0}{R}\right)^3\right]\frac{1-2\nu_1}{1-\nu_1}, \tag{A5}$$

$$u_{out}(r) = \varepsilon_0 r_0^3\left[\left(\frac{r}{R}\right)^3(2-4\nu_2)+1+\nu_2\right]\frac{1}{3r^2(-1+\nu_2)}, \tag{A6}$$

where $\nu_{1,(2)}$ are the Poisson ratios for the core (shell).

Then, the diagonal components of the strain tensor in spherical coordinate reads

$$\varepsilon_{rr} = \partial_r u(r), \ \varepsilon_{\theta\theta} = \varepsilon_{\varphi\varphi} = u(r)/r, \tag{A7}$$

and the hydrostatic strain is $\varepsilon_{hyd} = \varepsilon_{rr} + \varepsilon_{\theta\theta} + \varepsilon_{\varphi\varphi}$.

Thus, for the core we obtain

$$\varepsilon_{1rr} = \varepsilon_{1\theta\theta} = \varepsilon_{1\varphi\varphi} = \frac{2}{3}\varepsilon_0\left[1-\left(\frac{r_0}{R}\right)^3\right]\frac{1-2\nu_1}{1-\nu_1}, \tag{A8}$$

$$\varepsilon_{1hyd} = 2\varepsilon_0\left[1-\left(\frac{r_0}{R}\right)^3\right]\frac{1-2\nu_1}{1-\nu_1}, \tag{A9}$$

and for the shell

$$\varepsilon_{2rr} = -\frac{2\varepsilon_0}{3}\left(\frac{r_0}{r}\right)^3\left[\left(\frac{r}{R}\right)^3(-1+2\nu_2)+1+\nu_2\right]\frac{1}{(-1+\nu_2)}, \tag{A10}$$

$$\varepsilon_{2\theta\theta} = \varepsilon_{2\varphi\varphi} = \frac{\varepsilon_0}{3}\left(\frac{r_0}{r}\right)^3\left[\left(\frac{r}{R}\right)^3(2-4\nu_2)+1+\nu_2\right]\frac{1}{(-1+\nu_2)}, \tag{A11}$$

$$\varepsilon_{2hyd} = -\frac{2\varepsilon_0}{3}\left(\frac{r_0}{R}\right)^3\frac{1-2\nu_2}{1-\nu_2}. \tag{A12}$$

The Cartesian components are obtained with transformation relations between spherical and Cartesian tensor components[27]

$$\begin{cases} \varepsilon_{xx} = \sin^2\theta\cos^2\varphi\varepsilon_{rr} + \cos^2\theta\cos^2\varphi\varepsilon_{\theta\theta} + \sin^2\varphi\varepsilon_{\varphi\varphi} \\ \varepsilon_{yy} = \sin^2\theta\sin^2\varphi\varepsilon_{rr} + \cos^2\theta\sin^2\varphi\varepsilon_{\theta\theta} + \sin^2\varphi\varepsilon_{\varphi\varphi} \\ \varepsilon_{zz} = \cos^2\theta\varepsilon_{rr} + \sin^2\theta\sin^2\varphi\varepsilon_{\theta\theta} \\ \varepsilon_{xy} = \left(\sin^2\theta\varepsilon_{rr} + \cos^2\theta\varepsilon_{\theta\theta} - \varepsilon_{\varphi\varphi}\right)\sin 2\varphi \\ \varepsilon_{xz} = \left(\varepsilon_{rr} - \varepsilon_{\theta\theta}\right)\sin 2\theta\cos\varphi \\ \varepsilon_{yz} = \left(\varepsilon_{rr} - \varepsilon_{\theta\theta}\right)\sin 2\theta\sin\varphi \end{cases} . \tag{A13}$$

The resulting shear strain as shown by the above equation induces bands splitting. The splitting in some directions of the uniaxial strain has a linear dependence of the product between the Cartesian shear strain components or the difference $\varepsilon_{xx}-\varepsilon_{yy}$ [15] and the shear deformation potentials. According to Eq. (A11), we obtain $\varepsilon_{1xx}-\varepsilon_{1yy}=0$ , and $\varepsilon_{1xy}=\varepsilon_{1xz}=\varepsilon_{1yz}=0$ , and consequently, by neglecting the thin shell contribution to the splitting (for which $\varepsilon_{2xx}\neq\varepsilon_{2yy}$ , and $\varepsilon_{2xy},\varepsilon_{2xz},\varepsilon_{2yz}\neq 0$ ), we can approximate the band lineups without taking into account the splitting of the conduction or valence band by the shear strain.

## Appendix B: Heavy and light hole effective mass

Within the $\mathbf{k}\cdot\mathbf{p}$ theory, the heavy and light hole effective masses are given by[28]

$$m^{hh,(lh)}=\frac{m_0}{\gamma_1-(+)2\gamma}, \tag{B1}$$

with

$$\gamma=(1-\zeta)\gamma_2+\zeta\gamma_3,\quad \zeta(\theta,\varphi)=\sin^2\theta\left\{3-\frac{3}{8}\sin^2\theta[7+\cos(4\varphi)]\right\}, \tag{B2}$$

where $\theta,\ \varphi$ are the spherical coordinates of the direction perpendicular to the growth plan with respect to the Cartesian axes of the main crystallographic directions, and $\gamma_1,\ \gamma_2,\ \gamma_3$ are the Luttinger parameters. Then, to obtain the spherical part of the heavy-hole and light-hole masses assumed by the parabolic dispersion of the one-band model, we average over the solid angle the quantity $\zeta(\theta,\varphi)$

$$\bar{\zeta}=\frac{1}{4\pi}\int_0^{\pi}\int_0^{2\pi}d\theta d\varphi\sin\theta\zeta(\theta,\varphi)=\frac{3}{5}. \tag{B3}$$

Then, with the replacement $\gamma\to\bar{\gamma}=(1-\bar{\zeta})\gamma_2+\bar{\zeta}\gamma_3$ in Eq. (B1), we find Eq. (8), which recovers the expressions of Baldereschi and Lipari[29] of the heavy and light holes.

## Appendix C: Oscillator Strength

For the sake of self-consistency, we present derivation of the oscillator strength expressed by Eq. (10). Thus, we start with[30]

$$f_{ij}=\frac{2}{m_0\hbar\omega_{ij}}\left|\langle\Psi_i|\mathbf{e}\cdot\mathbf{p}|\Psi_j\rangle\right|^2=\frac{2}{m_0\hbar\omega_{ij}}\left|\langle\Psi_i|\mathbf{e}\cdot\frac{\hbar}{i}\nabla|\Psi_j\rangle\right|^2, \tag{C1}$$

where the total wave vector is the product of the envelope wave vector and cell Bloch wave vector, that is, $|\Psi_{nLm\alpha}\rangle = |\psi_{nLm}\rangle |u_\alpha\rangle$ ($\alpha = v, c$ for VB, CB, respectively). Then

$$\langle \Psi_i | \mathbf{e}\cdot\frac{\hbar}{i}\nabla | \Psi_j \rangle = \int_V \left( u_c^* u_v \psi_{nLm}^* \mathbf{e}\cdot\frac{\hbar}{i}\nabla \psi_{n'L'm'} + \psi_{nLm}^* \psi_{n'L'm'} u_c^* \mathbf{e}\cdot\frac{\hbar}{i}\nabla u_v \right) d\mathbf{r}, \quad \text{(C2)}$$

where $V$ is the volume of the QD. By making use of the slow spatial variation of the envelope wave function over regions of the unit cell size and the orthonormality of the Bloch cell wave functions, the above integral can be written as follows.

$$\begin{aligned} \langle \Psi_i | \mathbf{e}\cdot\frac{\hbar}{i}\nabla | \Psi_j \rangle &= \frac{\hbar}{i}\int_V \left( u_c^*(\mathbf{r}) u_v(\mathbf{r}) \psi_{nLm}^*(\mathbf{r}) \nabla \psi_{n'L'm'}(\mathbf{r}) + \psi_{nLm}^*(\mathbf{r}) \psi_{n'L'm'}(\mathbf{r}) u_c^*(\mathbf{r}) \nabla u_v(\mathbf{r}) \right) d\mathbf{r} \\ &= \int_V \psi_{nLm}^*(\mathbf{R}) \psi_{n'L'm'}(\mathbf{R}) d\mathbf{R} \int_\Omega u_c^*(\mathbf{r}) \mathbf{p}\, u_v(\mathbf{r}) d\mathbf{r} \equiv \langle \psi_{nLm}^* | \psi_{n'L'm'} \rangle \mathbf{p}_{cv} \end{aligned}, \quad \text{(C3)}$$

where $\Omega$ is the volume of the unit cell, $V$ is the hetero-structure volume, the capital $\mathbf{R}$ suggests an integration over coarse-grained unit cells space and $\int_\Omega u_c^*(\mathbf{r}) \frac{\hbar}{i} \nabla u_v(\mathbf{r}) d\mathbf{r} = \mathbf{p}_{cv}$ is the Bloch optical matrix element. By introducing the Kane momentum matrix element, $P = -i(\hbar/m_0)\langle s|p_z|z\rangle = -i(\hbar/m_0)\langle u_c|p_z|u_v\rangle = -i(\hbar/m_0)p_{cvz}$, with $E_P = 2m_0|P|^2/\hbar^2$, and considering the polarization unit vector, $\mathbf{e}$, parallel to the quantization axis, $z$, for example, the oscillator strength reads

$$f_{ij} = \frac{E_P}{\hbar\omega_{ij}} \left| \langle \psi_{nLm}^* | \psi_{n'L'm'} \rangle \right|^2 .$$

## References

[1] M. Bruchez, M. Moronne, P. Gin, S. Weiss, and A. P. Alivisatos, Science **281**, 2013 (1998); W. C. Chan and S. Nie, Science **281**, 2016 (1998).

[2] X. Michalet, F.F. Pinaud, L.A. Bentolila, J.M. Tsay, S. Doose, J.J. Li, G. Sundaresan, A.M. Wu, S.S. Gambhir, and S. Weiss, Science **307**, 538 (2005); X. Gao, Y. Cui, R.M. Levenson, L.W. Chung, and S. Nie, Nat. Biotechnol. **22**, 969 (2004); W. Cai, D.W. Shin, K. Chen, O. Gheysens, Q. Cao, S.X. Wang, S.S. Gambhir, and X. Chen, Nano Lett. **6**, 669 (2006).

[3] L. Shi, B. Hernandez, and M. Selke, J. Am. Chem. Soc. **128**, 6278 (2006).

[4] H. Zhu, N. Song, and T. Lian, J. Am. Chem. Soc. **133**, 8762 (2011), and the references therein.

[5] S. Pokrant and K.B. Whaley, Eur. Phys. J. D **6**, 255 (1999).

[6] S.J. Sun and Y.-C. Chang, Phys. Rev. B **62**, 13631 (2000).

[7] L. He, G. Bester, and A. Zunger, Phys. Rev. B **70**, 235316 (2004).

[8] J. Li and L. Wang, Appl. Phys. Lett. **84**, 3648 (2004).

[9] D. Schooss, A. Mews, A. Eychmüller, and H. Weller, Phys. Rev. B **49**, 17072(1994).

[10] G.A. Baraff and D. Gershoni, Phys. Rev. B **43**, 4011 (1991); D. Gershoni, C.H. Henry and G.A. Baraff, IEEE J. Quantum. Elect. **9**, 2433 (1993).

[11] P.C. Sercel and K. J. Vahala, Phys. Rev. B **42** 3690 (1990).

[12] E.P. Pokatilov, V.A. Fonoberov, V.M. Fomin, J.T. Devreese, Phys. Rev. B **64**, 245328 (2001).

[13] Y. Rajakararunnayake, R.H. Miles, G.Y. Wu, and T.C. McGill, Phys. Rev. B **37**, 10212 (1988).

[14] L.C. Lew Yan Voon and M. Willatzen, *The* $\mathbf{k}\cdot\mathbf{p}$ *Method, Electronic Properties of Semiconductors* (Springer-Verlag Berlin Heidelberg 2009) p.282.

[15] J. H. Davies, J. Appl. Phys. **84**,1358 (1998).

[16] C.G. Van de Walle, Phys. Rev. B **39**, 1871 (1989).

[17] O. Zakharov, A. Rubio, X. Blasé, M.L. Cohen, and S.G. Louie, Phys. Rev. B **50**, 10780 (1994).

[18] S. S. Lo, T. Mirkovic , C.-H. Chuang , C. Burda , and G. D. Scholes., Adv. Mater. **XX**, 1, (2010).

[19] P. Lawaetz, Phys. Rev. B 4, 3460 (1971).

[20] Singh J, *Physics of Semiconductors and Their Heterostructures* (McGraw-Hill, 1993).

[21] J. Bang, J. Park, J.H. Lee, N. Won, J. Nam, J. Lim, B.Y. Chang, H.J. Lee, B. Chon, J. Shin, J.B. Park, J.H. Choi, K. Cho, S.M. Park, T. Joo, and S. Kim, Chem. Mater. **22**, 233 (2010).

[22] A.M. Smith, A. M. Mohs, and S. Nie, Nat. Nanotechnol. **4**, 56 (2009).

[23] L. E. Brus, J. Chem. Phys. **79**, 5566 (1983).

[24] Akram K.S. Aqili, Zulfiqar Ali, Asghari Maqsood, Appl. Surf. Sci. **167**, 1 (2000).

[25] A. S. Saada, *Elasticity: Theory and Applications* (Pergamon New York, 1974).

[26] J. Rockenberger et al., J. Chem. Phys. **108**, 7807 (1998).

[27] See, e.g., T. Takagahara, J. Lum. **70** 129 (1996).

[28] R. Winkler, Spin–Orbit Coupling Effects in Two-Dimensional Electron and Hole Systems, Springer-Verlag Berlin Heidelberg 2003, section 4.5.1.

[29] A. Baldereschi and N. O. Lipari, Phys. Rev. B **8**, 2697 (1973).

[30] J. H. Davies, *The Physics of Low-Dimensional Semiconductors: An Introduction*, (Cambridge University Press, Cambridge, 1996), p.318.